\documentstyle[aps]{revtex}

\begin{document}
\author{R.V. TUMANIAN\\Yerevan Physics Institute,Armenia}
\title{RESONANT LASER COOLING of RELATIVISTIC CHARGES BEAMS}
\maketitle
\begin{abstract}

In work is considered average transverse dynamics of an electron beam
in the autoresonant laser. It is shown, that in approach of the given external
electromagnetic wave (small gain free electron laser) transverse emittance
of a beam of the charged particles decreases. This for the first time found out 
effect can be used for cooling beams of various accelerators of the charged
particles. In the field of particle energy about 100 in the mass energy units, 
beam energy losses are negligible. This method of charge beams cooling 
may be applied to electron, muon and various heavy particle beams.
\end{abstract}
\begin{flushleft}
\begin{twocolumn}
Enough methods of the charged particle accelerator beam cooling based on 
tested and used on various accelerators physical principles now are known: 
radiating [1], electron [2], stochastic [3] and Doppler [4] cooling. However they 
by the opportunities do not meet more modern requirements. Necessity for 
improvement of various parameters of these methods has resulted last years 
in consideration of new beam cooling methods from among which it is necessary 
to note autoresonant [5], undulator [6], ionization [7], laser [8] and stimulated 
radiation or Doppler [9] cooling. The special attention is given to cooling by 
backward Compton scattering of free electrons on photons [10-12]. It is known, 
however, that the stimulated scattering under other equal conditions has greater 
cross-section [ 9,13 ], than those of free particles. Therefore, as against the 
above-stated ways of cooling, in the given work the opportunity of transverse 
cooling of a beam of relativistic charges is considered at the stimulated or 
resonant interaction of particles with a flat electromagnetic wave (photons). 
For creation of conditions of resonant bunch - wave interaction is entered the 
homogeneous magnetic field directed along a direction of movement of a beam.
The magnetic field creates density correlations of a beam current on the laser 
wavelength that increases bunch - wave interaction and makes it selective.
Interaction moving on a periodic trajectory electrons with a flat monochromatic 
electromagnetic wave is one of possible ways of amplification of an 
electromagnetic wave in the free electron laser. Offered in [14] autoresonant 
free electron laser, based on interaction of the charged particle with the 
electromagnetic wave extending along a homogeneous magnetic field 
$B_0 \parallel z$, is characterized by a detuning constancy from an exact 
resonance at charged particle energy change. In [15] shown, that the detuning 
constancy connected to existence of motion integral
\[
I = \gamma  - P_z 
\]
takes place only in approach of the given external wave. Here $\gamma$ is the 
full energy, and $P$ momentum of a particle along a magnetic field (in terms 
of m=c=1). In the present work influence of autoresonant interaction of a particle 
with a wave on a beam transverse emittance is examined. Taking into account 
connection $\gamma  = \sqrt {P_z^2  + P_ \bot ^2  + 1} $, from (1) it is easy to 
receive for a transverse kinematic momentum of a particle
\[
P_ \bot ^2  = 2I\gamma  - 1 - I^2 
\]
Differentiating this expression on time and averaging on all particles of a beam,
we shall receive the equation for change of an average square of a 
cross-kinematic momentum of a beam
\[
\frac{d}{{dt}}\langle P_ \bot ^2 \rangle  = 2\langle I\dot \gamma \rangle 
\]
As shown in works [14-16], trajectories of electrons are screw lines with radius 
varying along a way. As drift of the center of a Larmor circle absence and I > 0, 
interaction in a mode of the laser ($\dot \gamma  < 0$) means reduction of a 
beam transverse emittance, because of negativity of 
the right part in (2). In a laser mode the amplified wave carried away with itself 
a part of beam energy. It is very important to find what part of wave carried 
energy is from full beam energy and what part is from beam emittance, which 
means beam cooling (emittance reduction). As it is not difficult to receive from 
(1), relative speed $\dot P_z /P_z $ of a longitudinal momentum change 
$2P_ \bot ^2 /(1 + P_ \bot ^2 )$ times less than relative speed 
$\dot P_ \bot  /P_ \bot  $ of a cross-kinematic momentum change. Therefore 
emittance reduction will be effective for beams with 
$\langle P_ \bot ^2 \rangle  \ll 1$, when full energy losses are small. But in 
practical applications it is possible also to compensate beam energy losses 
by external accelerating elements. In approach of infinitesimal changes of a 
cross-kinematic momentum 
at interaction with a wave, when motion in a cross plane is determined by a 
magnetic field $B_0^{} $, for $\dot \gamma $ in a field of the circular-polarized 
wave with frequency ( and amplitude Å it is easy to receive 	
\[
\dot \gamma  = \xi \omega \frac{{P_ \bot  }}{\gamma }\cos \varphi 
\]
where $\xi  = {\raise0.7ex\hbox{${eE}$} \!\mathord{\left/
 {\vphantom {{eE} \omega }}\right.\kern-\nulldelimiterspace}
\!\lower0.7ex\hbox{$\omega $}}$ is the dimensionless amplitude of a wave 
electric field, and ( is a phase of the Larmor rotations, counted from a wave 
phase in a point of a presence of a particle. In used by us approach it is easy 
to receive the equation for change of an average longitudinal momentum or 
energy of a beam
\[
\left\langle {\dot p} \right\rangle  = \left\langle {\dot \gamma } \right\rangle  = \left\langle {\xi \omega \frac{{p_ \bot  }}{\gamma }\cos \varphi } \right\rangle 
\]
Averaging on all particles of a beam in the equation (2) allows instead of a 
detailed trajectory of one particle in the given external fields to investigate 
average on a beam of size, using the kinetic approach. At presence of an 
electromagnetic wave with $\vec E,\vec B \sim e^{i(\omega t - kz)} $
which we shall count small in comparison by a magnetic field $\vec B_0 $, 
in a beam there are correlations of density which result in modulation of a 
beam on ( and accordingly, to a nonzero right part (2). For calculation of this 
effect we shall present function of distribution of a beam as  [17]
$f = f_0  + \delta f$, where $f_0 $ is the basic equilibrium function of distribution, 
and $\left| {\delta f} \right| \ll f_0 $ is the small amendment to $f_0 $, cause by 
a wave. Near to a resonance it is convenient to present the amendment as 
Fourier series from a phase 
$\varphi :\delta f = \sum\limits_{}^{} {g_s (P_ \bot  ,P_z )e^{is\varphi } } $. 
Substituting in the kinetic equation and neglecting members of the second 
order, we shall receive the following decision for factors
\[
g_s  = \frac{{Q_s }}{{i(s + \alpha )}};Q_s  = \frac{1}{{2\pi }}\int\limits_0^{2\pi } {d\tau e^{ - is\tau } Q(P_z ,P_ \bot  ,\tau );Q = \frac{e}{{\omega _B }}\frac{{\partial f_0 }}{{\partial \vec P}}(\vec E + \frac{1}{\omega }[\vec v[\vec k\vec E]])} 
\]
where $\alpha  = {\raise0.7ex\hbox{${(kv_z  - \omega )}$} \!\mathord{\left/
 {\vphantom {{(kv_z  - \omega )} {\omega _B }}}\right.\kern-\nulldelimiterspace}
\!\lower0.7ex\hbox{${\omega _B }$}}$,  
$\omega _B  = {\raise0.7ex\hbox{${eB_0 }$} \!\mathord{\left/
 {\vphantom {{eB_0 } \gamma }}\right.\kern-\nulldelimiterspace}
\!\lower0.7ex\hbox{$\gamma $}}$  is the Larmore frequency,$\tau$ is the phase 
of integration and used $\omega \vec B = [\vec k\vec E]$
connection for a field of a wave. Near to a simple cyclotron resonance the basic 
member in Four-decomposition will be a member with 
s=1. For Gaussian distribution function on a cross momentum 
\[
f_0  = \frac{1}{{\pi \left\langle {P_ \bot ^2 } \right\rangle }}e^{ - {\raise0.7ex\hbox{${P_ \bot ^2 }$} \!\mathord{\left/
 {\vphantom {{P_ \bot ^2 } {\left\langle {P_ \bot ^2 } \right\rangle }}}\right.\kern-\nulldelimiterspace}
\!\lower0.7ex\hbox{${\left\langle {P_ \bot ^2 } \right\rangle }$}}} \delta (P_z  - P_0 )
\]
with the account $I \approx {{(1 + P_ \bot ^2 )} \mathord{\left/
 {\vphantom {{(1 + P_ \bot ^2 )} {2P_0 }}} \right.
 \kern-\nulldelimiterspace} {2P_0 }}$ in a relativistic case $P_0  \gg 1 + P_ \bot ^2 $ 
we shall receive
\[
\delta f =  - \xi \frac{{P_ \bot  (1 + P_ \bot ^2 )}}{{\left\langle {P_ \bot ^2 } \right\rangle i(\Delta _\parallel   - P_ \bot ^2 )}}e^{i\varphi } f_0 
\]
 where
$\Delta _\parallel   = 2\frac{\Omega }{\omega }P_0  - 1;\Omega  = \omega _B \gamma $. 
Approximation of a longitudinal - monochromatic beam in the $f_0 $ as it is 
easy to estimate from resonant term $(1 + \alpha )^{ - 1} $, is fair, if the relative 
spread on a longitudinal momentum $(P_z  - P_o )/P_0 $ is much less
$\left\langle {P_ \bot ^2 } \right\rangle $. At averaging on f there is a usual pole 
in such cases in a point  $p_ \bot ^2  = \Delta _\parallel  $, caused by a 
resonant multiplier. By a rule of detour of Landau poles  (replacement 
$\omega  \to \omega  + i0$) we receive for integral interesting us [17]: 

\[
\int\limits_0^\infty  {\frac{{f(z)}}{{\Delta _\parallel   - z}}} dz = \int {\frac{{f(z)}}{{\Delta _\parallel   - z}}dz}  + i\pi \delta (\Delta _\parallel   - z)
\]
where $z = p_ \bot ^2 $, and the right integral is meant in sense of a principal 
value. As the real part of the equation (2), has physical sense only at 
$\Delta _\parallel   \le 0$ if the right part only imaginary, speed of change of a 
cross momentum equal to zero. And we receive: 

\[
\frac{d}{{dt}}\left\langle {p_ \bot ^2 } \right\rangle  =  - \xi \frac{{\pi \omega }}{{2p_0^2 \left\langle {p_ \bot ^2 } \right\rangle }}\Delta _\parallel  (1 + \Delta _\parallel  )e^{ - {{\Delta _\parallel  } \mathord{\left/
 {\vphantom {{\Delta _\parallel  } {\left\langle {p_ \bot ^2 } \right\rangle }}} \right.
 \kern-\nulldelimiterspace} {\left\langle {p_ \bot ^2 } \right\rangle }}} 
\]

Maximum of the right part in a case $\left\langle {p_ \bot ^2 } \right\rangle  \ll 1$
($p_0  \approx const$) is achieved in a point
$\Delta _\parallel   = \left\langle {p_ \bot ^2 } \right\rangle $, and its solution is
\[
\left\langle {p_ \bot ^2 } \right\rangle  = (\left\langle {p_{ \bot 0}^2 } \right\rangle ^2  - \frac{{2\pi ^2 }}{e}\xi ^2 \frac{z}{{2\lambda p_0^2 }})^{1/2} 
\]
where $P_{\bot 0}$ is the initial value of an average square of a cross momentum of a beam, 
($\lambda$ is a length of a wave, and z =ct is a length of a beam way. If 
$\Delta _\parallel   = const$ 
and much less $\left\langle {p_ \bot ^2 } \right\rangle $, we shall receive: 
\[
\left\langle {p_ \bot ^2 } \right\rangle  = (\left\langle {p_{ \bot 0}^2 } \right\rangle ^3  - \xi ^2 \frac{{6\pi ^2 z}}{{2\lambda p_0^2 }}\Delta _\parallel  )^{1/3} 
\]

Decreasing of a beam  cross emittance at autoresonant bunch - wave interaction 
must be significant as well for a bunch modulated on $\phi$ which function of 
distribution we shall choose in such kind: 
\[
f(p_ \bot  ,p_z ,\varphi ) = f_0 (1 - \varepsilon \cos \varphi )
\]
where $0 < \varepsilon  \le 1$ is the depth of modulation, and $f_0 $ is defined 
from (5). After averaging (2) on this function we shall receive: 
\[
\frac{d}{{dt}}\left\langle {p_ \bot  } \right\rangle  =  - \varepsilon \xi \frac{\omega }{{2p_0^2 }}(1 + \frac{6}{\pi }\left\langle {p_ \bot  } \right\rangle ^2 ),
\]
where
$\left\langle {p_ \bot  } \right\rangle  = \frac{{\sqrt \pi  }}{2}\sqrt {\left\langle {p_ \bot ^2 } \right\rangle } $
connection is taken into account. From here at 
$\left\langle {p_ \bot ^2 } \right\rangle  \ll 1$ ($p_0  \approx const$), we shall 
receive 
	
\[
\left\langle {p_ \bot  } \right\rangle  = p_{ \bot 0}  - \varepsilon \xi \pi \frac{z}{{2\lambda p_0^2 }}
\]
The condition of small changes of the cross momentum, used at a conclusion 
of all formulas, means actually restriction on passed way length: 
\[
1 \le \frac{z}{{2\lambda p_0^2 }} \ll A
\]
where
$A = \frac{{e\left\langle {p_{ \bot 0}^2 } \right\rangle ^2 }}{{2\pi ^2 \xi ^2 }};\frac{{\left\langle {p_{ \bot 0}^2 } \right\rangle ^3 }}{{2\pi ^2 \xi ^2 \Delta _\parallel  }};\frac{{p_{ \bot 0} }}{{\varepsilon \xi \pi }}$
for formulas (8), (9) and (12) respectively. And the left inequality is consequence 
to approach of adiabatic switching of a wave field. As in our case the field is 
switched instantly the approach adiabatic slow engaging of a wave field is fair 
at times greater than relaxation time. That is expression (6) for fair if the beam 
has passed a way greater than $2 p_0^2\lambda$. We shall emphasize that 
the condition (13) follows from a method of calculation. The equation (2) fairly in 
approach of the given external field when Leangmure frequency of a beam is 
much less than frequency of the laser [15]. It specifies connection between 
speed of transverse momentum change and speed of a beam energy change. 
For a beam with$P_0=100$ and divergence $\vartheta=10^{-3}$ the laser with 
$\xi=10^{-2}$ the length of a way on which transverse emittance change to 100 
makes 80 cm, thus length of a relaxation about 20cm. Thus, autoresonant 
cooling of beams of relativistic charges is much effective and faster than known 
and listed above methods of cooling. It is very effective for beams with the 
Lorenz-factor about 100, but is applicable also to GeV particles and 
for achievement comparable to \cite{10,11} rates of cooling sources of 
electromagnetic radiation, in this case masers, with rather low and more real 
capacities are required. We shall notice that intensity of cooling depends on 
parameter $\xi$, which is easy for making about unity in long-wave sources of 
radiation, for example in masers. Thus it is much less than loss of energy of 
a beam. If to add to this an opportunity of application of the longitudinal and - or 
cross-section non-uniform magnetic field that enables scanning in case of a 
beam with the big divergence, and also selective cooling advantage of this 
method becomes obvious. Hence, autoresonant cooling, that is laser cooling 
of beams of charges in a longitudinal magnetic field has a number of advantages 
even in comparison with the newest laser \cite{10, 11} and undulator cooling 
methods, as 
because of a dependence of cooling speed from a cross momentum of particles 
and a beam as a whole, and in connection with formation of correlations of 
currents of a beam and ample opportunities of their management by 
heterogeneity of a magnetic field for increase of efficiency and selectivity of 
cooling. In this sense it is comparable to the stimulated cooling methods. This 
cooling, if applied, can to increase significantly the luminosity of the storage 
ring and synchrotron beams, due to decrease of the beam sizes. Under certain 
conditions autoresonant cooling can succesfully applied to linac beams too 
and give a possibility to achieve  highest density and luminosity beams with 
small sizes and momentum spread.

\end{twocolumn}

\end{flushleft}
\end{document}